\documentclass[11pt]{article}

\usepackage{graphicx}
\usepackage{amssymb}
\usepackage{amsmath}

\begin{document}

\begin{center}
\textbf{\begin{LARGE}The proton gyromagnetic g-factor:\end{LARGE}\\
\begin{Large}An electromagnetic model\end{Large}}
\end{center}

\begin{small}
\begin{center}
\textbf{G. Sardin}
\end{center}

\begin{center}
Universitat de Barcelona - Facultat de Fisica\\gsardin@ub.edu
\end{center}
\end{small}

\begin{abstract}

\begin{footnotesize}
So far, the Standard Model of Elementary Particles has not succeeded getting a trustworthy account of the proton spin, which remains an enigma. This hindrance is known as the proton spin crisis, owing to the experimental evidence already from 1988 suggesting that little or none of the proton's spin would come from the spin of the quarks. This prompted theorists to a flood of guessworks about the proton's spin. Since it remains unsolved, in the framework of new physics an exploratory approach based on a novel paradigm is proposed, which brings a renewed access to this challenge, through its reciprocal relationship with the g-factor.
\\\\
The Orbital Model of Elementary Particles allows deepening the physical significance of the gyromagnetic g-factor by correlating it to structural and inner dynamical parameters. This new approach provides a further insight to the correlation between magnetic moment and mass of elementary particles, through the relationship between the electromagnetic radius and wavelength of their structure. The deBroglie and Compton wavelengths can be used equivalently. The structure of elementary particles can be efficiently approached by considering them to be defined by two inner dynamics, a rotation and an oscillation of the electric charge tracing the structuring orbital. To the rotation of the electric charge is associated the magnetic moment, which depends on the orbital radius, and to its oscillation is associated the mass, which depends on the structure wavelength. The relationship between the two dynamics, gyratory and oscillatory, defines the quantization of the structuring orbital. The correlation between the magnetic moment and the mass of different particles suggests that they could be weakly coupled. The quantization of the structure radius and wavelength fixes their relationship, which in turn determines the relationship between $\mu$ and m. Here its application is addressed to the proton. Its g-factor is related to its structural state. Its reduced structure wavelength and electromagnetic radius differ slightly and their ratio $r/\bar{\lambda}$ fixes the value of the g-factor.
\\\\
\textbf{Pacs:} 12.90.+b (Miscellaneous theoretical ideas and models)
\\14.20.Dh (Properties of protons) 		
41.20.-q (Applied classical electromagnetism) \\
\end{footnotesize}
\end{abstract}

\section{Introduction}

Up to date, the Standard Model of Elementary Particles has not succeeded getting a reliable quantitative account of the proton spin, which remains an unsolved puzzle. In effect, in 1988, physicists were shocked to find experimental evidence suggesting that very little or perhaps none of the proton's spin comes from the spin of the quarks thought to make up the proton. They called this apparent inconsistency the proton spin crisis, which twenty years later is still current (1-12). This incited theorists to a torrent of guessworks  about the proton's spin, which is the source of its magnetic moment and associated g-factor. 
\\\\
Based on new physics, the Orbital Model of elementary particles [13,14], brings a refreshed access to the proton spin crisis. In particular, it allows a novel approach to the gyromagnetic g-factor by relating it to structural properties defined by the structure wavelength and radius, as well as dynamical by way of the gyratory and oscillatory frequencies, instead of expressing just the relationship between magnetic moment and spin, or equivalently between theoretical and experimental magnetic moments. Likewise, this model provides a straightforward apprehension of the correlation between the magnetic moment and the mass of elementary particles, through the relationship between their structure wavelength: $\bar{\lambda} = \hbar/m v$, and radius: $r = 2\mu/e v$. 
\\\\
Let us briefly sketch the premises of the orbital model of the structure of elementary particles. These are primarily regarded as being embodied by orbitals, formed by a single or a pair of integer electric charges. The orbital concept is basically extrapolated from the atomic structure, shaped by orbitals, however those of elementary particles differ in being nucleusless and self-confined [13,14]. To the inherent inner dynamics of the structure orbital is associated the magnetic moment, defined by the radius and gyratory speed, as well as the mass which derives from the structure wavelength. The diversity of elementary particles, along with their different mass and magnetic moment, arises from the variety of quantum states $\left|\psi\right\rangle$ that may acquire their structure, which is conformed by a unique structural scheme. The orbital of singly charged particles is shaped by a single electric charge, while that of neutral particles is shaped by a pair of electric charges of opposite sign. A procedure for the quantization of the proton mass has been proposed [14].
\\\\
Let us point out that the knowledge of the actual shape of the structuring orbital is not a compulsory requirement, in the same way as from the Bohr model energy levels are derived with a fairly high accuracy without the accurate knowledge of the exact shape of the atomic orbitals, which are reduced to just a circular outline. Somewhat surprisingly, the g-factor is fairly uncaring of the exact quantum shape of the structuring orbital and a semi-classical handling is quite effective, just as for the atomic structure. We have found that most properties of elementary particles can be treated semi-classically, procedure which presents the advantage of being conceptually simple and allowing not to loose sight of the physical depiction of their structure. 
\\\\
In other words, the orbital structure of elementary particles can be efficiently approached by considering them to be basically defined by two inner dynamics, a rotation and an oscillation of the structuring electric charge. To the rotation of the electric charge is associated the magnetic moment, which depends on the orbital radius, and to its oscillation is associated the mass, which depends on the structure wavelength. These two dynamics, gyratory and oscillatory, define the quantization of the structuring orbital, leading to its multiplicity. The correlation between the magnetic moment and the mass of elementary particles suggests some type of coupling, allowing different relationships. The variety of quantization of their structure leads to the different types of coupling. Specifically, the quantization of the orbital radius and the structure wavelength fixes their coupling, which in turn determines the relationship between $\mu$ and m. The ratio $r/\bar{\lambda}_{C}$ typifies the gyromagnetic g-factor. In this article only the proton g-factor is addressed. 

\section{Developments}
Elementary particles, regarded as orbital systems, can be quite appropriately approached by means of two essential parameters: the electromagnetic radius, to which is associated the magnetic moment, and the structure wavelength, to which is associated the mass. Let us first address a method to quantize the gyratory speed of the structural electric charge.

\subsection{Quantization procedure of the magnetic moment gyratory speed}

Since from the orbital model of the structure of elementary particles the gyratory speed of their vector electric charge appears to be fairly high, the Lorentz $\gamma$ factor has thus to be used. So, for mathematical handling let us apply $\gamma$ to the speed itself as expressed in the following formulation:
$v_{y} = v_{x}\gamma^{-1} = v_{x}\sqrt{1-v_{x}^{2}/c^{2}}$, in order to get a pondered value $v_{y}$ of the actual speed $v_{x}$. For $v_{x}$ varying from 0 to c, the graph (fig.1) shows a maximum pondered speed $v_{y} = c/2 = 1.49896\: 10^{8}$ m/s for the actual speed $v_{x} = c/\sqrt{2} = 2.11985\: 10^{8}$ m/s.

\begin{figure}[h!]
	\includegraphics[scale=0.48]{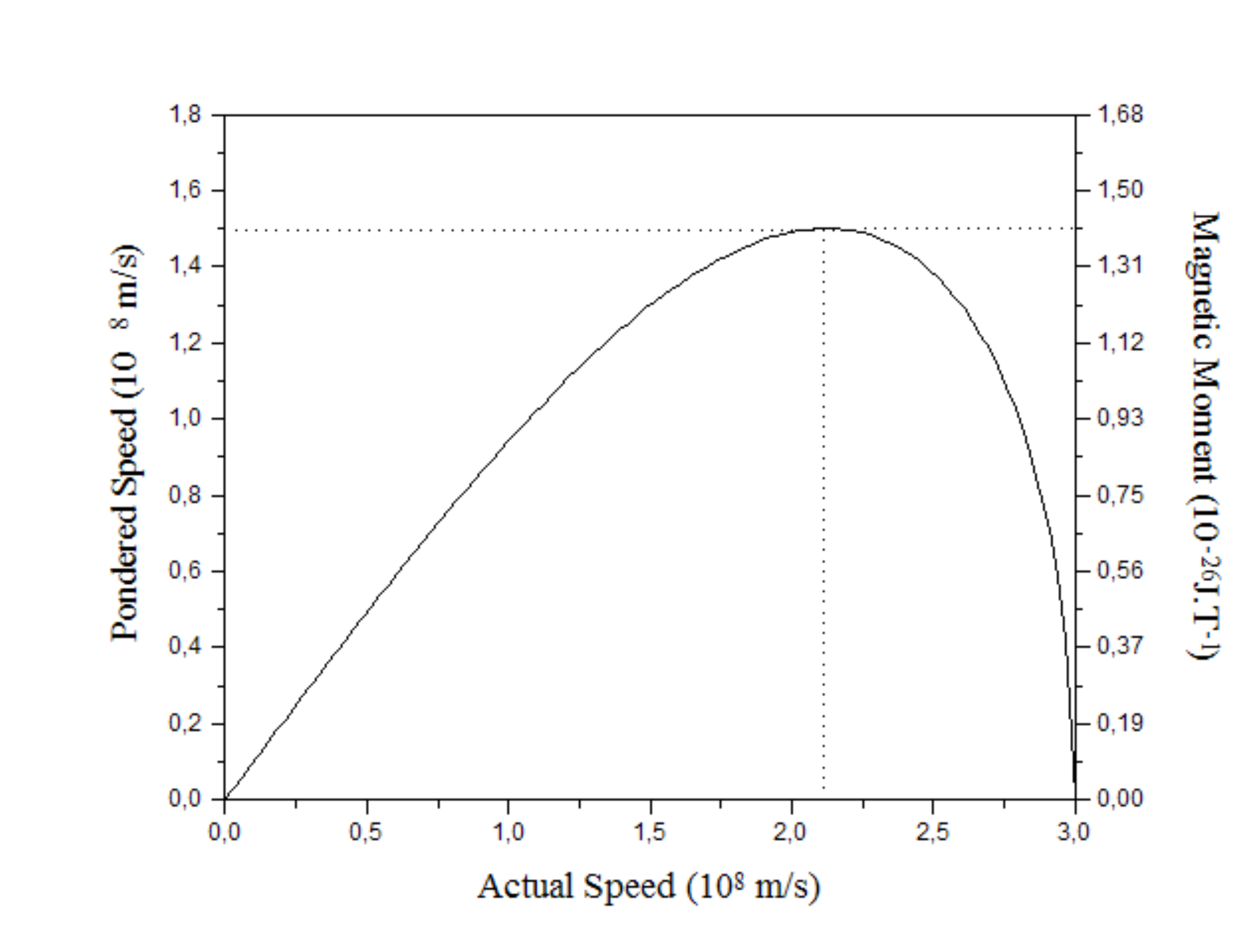}
	\caption{\textit{Quantization procedure of the gyratory speed: the maximum of the curve $\gamma^{-1} v$ is reached for the actual speed $v_{x} = c/\sqrt{2} = 2.11985\: 10^{8}$ m/s, and the pondered speed $v_{y} = c/2 = 1.49896\: 10^{8}$ m/s. Both speeds can be used equivalently and may be related to the deBroglie wavelength as well as to the Compton wavelength.}}
	\label{fig: 1}
\end{figure}

On physical grounds dealing with quantum orbital systems of femtometric size, the $\gamma$ factor is considered to express the fact that the gyratory speed of the vector electric charge actually looses efficiency in the production of magnetic moment as it increases, so its strength is not directly proportional to the speed. As speed increases it reaches a maximum at $2.11985\: 10^{8}$ m/s, and then decreases for still increasing speed, down to zero when reaching the speed of light (fig.1).
\\\\
The g-factor of the proton is derived using the structure wavelength, which can be defined by means of the deBroglie wavelength as well as the Compton wavelength. The deBroglie wavelength is related to $v$ ($\bar{\lambda}_{B} = \hbar/m_{p} v$), which is equal to the actual speed $v_{x} = c/\sqrt{2}$, so it can be expressed in function of c, i.e. $\bar{\lambda}_{B} = \hbar/m_{p} v_{x} = \hbar/(m_{p} c/\sqrt{2})$, while by definition the Compton wavelength is directly related with $c$ ($\bar{\lambda}_{C} = \hbar/m c$), so it can be related to the pondered speed which is equal to $v_{y} = \gamma^{-1} v_{x} = c/2$, and thus $\bar{\lambda}_{C} = 2\hbar/m_{p} v_{y}=\hbar/m_{p} c$. At this point, there are two possibilities, one consisting in using the actual speed $v_{x}$ and the other one in using the pondered speed $v_{y}$, whose ratio $v_{x}/v_{y} = \sqrt{2}$.
\\\\
In the first case the magnetic moment is defined by $v_{x}$, so: $\mu_{B} = \small\frac{1}{2} e v_{x} r_{B} = \small\frac{1}{2} e (c/\sqrt{2})r_{B}$ and in the second case it is defined by $v_{y}$, so: $\mu_{C} = \small\frac{1}{2}e v_{y} r_{C} = \small\frac{1}{2} e (c/2) r_{C}$.  Using the speed $v_{x}$ makes the magnetic moment to be ruled by the deBroglie radius $r_{B}$ and wavelength $\bar{\lambda}_{B}$, while using the speed $v_{y}$ makes it instead ruled by the Compton radius $r_{C}$ and wavelength $\bar{\lambda}_{C}$. Both cases can be used equivalently in regard to the formulation of the magnetic moment and the g-factor. Since $\bar{\lambda}_{B} = \hbar/m_{p}v_{x} = \hbar/(m_{p}c/\sqrt{2})$ and $\bar{\lambda}_{B} = \hbar/m_{p}c$, the ratio $\bar{\lambda}_{B}/\bar{\lambda}_{C} = \sqrt{2}$ = 1.4142. 

\subsection{Derivation of the proton $r_{B}$ and $r_{C}$ radii and their respective relation with its structure wavelengths $\bar{\lambda}_{B}$ and $\bar{\lambda}_{C}$}

Since the definition of the magnetic moment for a gyratory electric charge with a circular orbit is $\muµ = i A$, where: $i = e/\tau$, $\tau = 2\pi r/v$ and $A = \pi r^{2}$, its classical formulation is thus  $\mu = \small\frac{1}{2} e v r$. For femtometric systems such as elementary particles, the Lorentz scaling factor $\gamma$ has to be added to the classical formulation of the magnetic moment, since the speed of the structuring electric charge is high, and thus $\mu = \small\frac{1}{2} e (\gamma^{-1}v)r$. Let us point out that the $\gamma$ factor is here considered not to express any contraction of length and thus it is not applied to the radius [14], but to the gyratory speed itself, expressing the fact that speed looses efficiency in the production of the magnetic moment as it increases, so its strength is not directly proportional to the actual speed. Therefore, it is useful to differentiate between actual speed ($v$) and pondered speed ($\gamma^{-1}v$). At the curve maximum, the actual speed of the structuring electric charge is: $v_{x} = c/\sqrt{2}$, while the pondered speed reaches a maximum value that is lower: $v_{y} = c/2$, and decreases when $v_{x}$ further increases. In regard to the magnetic moment formulation both speeds can be used. 
\\\\
Still, the magnetic moment of elementary particles is quantized, taking so discrete values. The quantization factor corresponds to the relationship between the structure wavelength and radius, and may be expressed in terms of the deBroglie wavelength and radius, or equivalently in terms of the Compton wavelength and radius.
\\\\
Let us now apply the gyratory speed $v_{x}$ and $v_{y}$ to the expression of the proton magnetic moment, and from the curve let us select the maximum value of the magnetic moment, so:
\\\\
$\mu_{p} = \small\frac{1}{2} e v_{x} r_{B} = \small\frac{1}{2} e (c/\sqrt{2})r_{B}$ 
\\\\
or equivalently: 
\\\\
$\mu_{p} = \small\frac{1}{2} e v_{y} r_{C} = \small\frac{1}{2} e (c/2) r_{C}$
\\\\
As $v_{x}$ increases from 0 to $2.11985\: 10^{8}$ m/s, the magnetic moment $\mu_{p}$ increases, and starts decreasing for higher values of $v_{x}$, down to zero for $v_{x} = c$. The evolution of the magnetic moment reaches thus a maximum for $v_{x} = c/\sqrt{2}$, corresponding to the maximum value of the pondered speed $v_{y}$, which is then equal to c/2. Let us calculate the corresponding radius: 
\\\\
$r_{B} = \mu_{p} / [\small\frac{1}{2} e\; v_{x}] = \muµ_{p} / [\small\frac{1}{2} e\;(c/\sqrt{2})]$
\\\\
which gives: $r_{B} = 0.83065\: 10^{-15}$ m
\\\\
$r_{C} = \mu_{p} / [\small\frac{1}{2} e\; v_{y}] = \mu_{p} / [\small\frac{1}{2} e\; (c/2)]$
\\\\
which gives: $r_{C} = 1.17472\: 10^{-15}$ m
\\\\
where $\muµ_{p}$ is its experimental value. So, the applied method leads to a deBroglie radius of $0.83065\: 10^{-15}$ m and a Compton radius of $1.17472\: 10^{-15}$ m. Let us mention that the Compton radius is not the actual electromagnetic radius of the proton, which is instead set by the deBroglie radius since its associated speed $v_{x}$ is the actual speed of the carrier electric charge, while the Compton radius is associated to the pondered speed $v_{y}$, and so expresses the radius that would have the proton if the speed of the structuring electric charge would be c/2. Even though the Compton radius is virtual, its utility stands in allowing to correlate the magnetic moment to the full speed c and thus to the Compton wavelength as well. Let us stress that the pondered Compton radius is equal to the deBroglie radius:
\\\\
$r_{B} = r_{C} \gamma^{-1} = r_{C} \sqrt{1-v^{2}_{x}/c_{2}} = 0.83065\: 10^{-15}$ m
\\\\
where $\gamma$ plays here the role of a corrective factor, instead of expressing any actual physical contraction of length, due to the fact that the Compton radius as well as the Compton wavelength are derived using the speed c/2 and c respectively, instead of the actual speed $v_{x}$. So, the actual proton radius is directly provided by the deBroglie radius, but it can also be obtained from the pondered Compton radius, both being equal to 0.83 Fm, value to be compared to the experimental one of 0.85 Fm [12].
\\\\
Let us now calculate the reduced deBroglie $\bar{\lambda}_{B}$ and Compton $\bar{\lambda}_{C}$ wavelengths:
\\\\
$\bar{\lambda}_{B} = \hbar / m_{p} v_{x} = 1.0546\: 10^{-34} / [(1.6726\: 10^{-27})(2.1198\: 10^{8})]$
\\\\
$\bar{\lambda}_{B}  = 0.29742\: 10^{-15}$ m $\approx 0.30$ Fm
\\\\
$\bar{\lambda}_{C} = \hbar / m_{p} c = 1.0546\: 10^{-34} / [(1.6726\: 10^{-27})(2.9979\: 10^{8})]$
\\\\
$\bar{\lambda}_{C} = 0.21031\: 10^{-15}$ m $\approx 0.20$ Fm
\\\\
The reduced deBroglie ($\bar{\lambda}_{B} = \lambda_{B}/2\pi$) and Compton ($\bar{\lambda}_{C} = \lambda_{C}/2\pi$) wavelengths are innate magnitudes for mass at the quantum scale.
\\\\
Let us next assess the ratio between their respective radius and reduced wavelength:
\\\\
$\eta_{B} = r_{B} / \bar{\lambda}_{B} = 0.8306\: 10^{-15} / 0.2974\: 10^{-15} = 2.7928 = g/2$
\\\\
$\eta_{C} = r_{C} / \bar{\lambda}_{C} = 1.1747\: 10^{-15} / 0.21031\: 10^{-15} = 5.5857 = g$
\\\\
So, both values $\eta_{B}$ and $\eta_{C}$ of the gyro-oscillatory $\eta$-factor express the experimental gyromagnetic g-factor of the proton [5, 6]. The factor 2 between the structural parameters $\eta_{B}$ and $\eta_{C}$ arises from the fact that in deference to the definition of the Compton wavelength it has been defined with the speed c, instead of the pondered speed c/2 used. 
\\\\
Let us highlight that the value of 0.83 Fm found for the proton radius is the exclusive value that provides the proton g-factor, corresponding to a gyratory speed $c/\sqrt{2}$ of the electric charge, for which the resultant magnetic moment is maximum. Therefore, the described approach provides a cutting edge structural origin to the proton g-factor, expressing the ratio between radius and wavelength, which is a step more specific than the corrective g-factor of the ratio of the experimental value and the calculated one from $\mu = (e/m) S$, where S is the spin ($S = \hbar/2$). For the proton the gyro-oscillatory $\eta$-factor, which corresponds to the ratio $r/\bar{\lambda}$, expresses the gyromagnetic g-factor. Let us stress that $\eta$ is a structural factor since it relates two lengths ($r$ and $\bar{\lambda}$), while g relates the experimental and theoretical magnetic moments. 
\\\\
In a wider scope, the ratio $\eta = r/\bar{\lambda}$ of the two characteristic lengths, the structure radius and  wavelength, is equal to $\alpha^{-1}$ when the orbital radius is derived from the electrostatic potential ($V = -e^{2}/r$), but differs when the quantization of the orbital radius obeys to some other rule. For example, the electron magnetic moment is proportional to its Compton wavelength $\bar{\lambda}_{C}$ ($\mu_{e} = e\; c/2\; \bar{\lambda}_{C}$) and to its classical radius through the inverse fine-structure constant $\alpha^{-1}$  ($\mu_{e} = \alpha^{-1}\; e\; c/2\; r_{e}$), but that of the proton, when also related with its Compton wavelength $\bar{\lambda}_{C}$ and associated radius ($\mu_{p} = \small\frac{1}{2}\; e\; c/2\; r_{C}$), is proportional to its g-factor ($\mu_{p} = \small\frac{1}{2}\; g\; e\; c/2\; \bar{\lambda}_{C}$). So:
\\\\
$\mu_{p} = \small\frac{1}{2} \eta e (c/2) \bar{\lambda}_{C} = \small\frac{1}{2}(r_{C}/\bar{\lambda}_{C}) e (c/2) \bar{\lambda}_{C}$
\\\\
$\mu_{p} = \small\frac{1}{2} g e (c/2) \bar{\lambda}_{C} = 2.79284 e (c/2) \bar{\lambda}_{C}$
\\\\
and:
\\\\
$\mu_{e} = \eta^{-1} e (c/2) re = (r_{e} /\bar{\lambda}_{C})^{-1} e (c/2) r_{e}$ 
\\\\
$\mu_{e} = (\hbar c/e^{2}) e (c/2) r_{e} = \alpha^{-1} e (c/2) r_{e} = 137.036 e (c/2) r_{e}$
\\\\
So, while $r_{C}/\bar{\lambda}_{C} = g$ for the proton, $r_{e}/\bar{\lambda}_{C} = \alpha$ for the electron.

\subsection{Correlation between oscillatory and gyratory frequencies}

Besides the structural relationship of the $\eta$-factor between the orbital radius and wavelength, it can also be formulated in dynamical terms, i.e. rotation and oscillation. 
\\\\
The reduced deBroglie oscillatory frequency $\bar{\nu}_{B}$ is:
\\\\
$\bar{\nu}_{B} = v_{x}/\bar{\lambda}_{B} = 7.12743\: 10^{23}$ Hz
\\\\
and the reduced deBroglie gyratory frequency $\bar{\nu}_{g}$ is:
\\\\
$\bar{\nu}_{g} = v_{x}/r_{B} = c/\sqrt{2}r_{B} = 2.55203\: 10^{23}$ Hz
\\\\
So, their ratio $\eta_{B} = \nu(oscillatory) / \nu(gyratory)$ is:
\\\\
$\eta_{B} = \bar{\nu}_{B}/\bar{\nu}_{g}$ = 2.79285 = g/2
\\\\
and the proton magnetic moment expressed in terms of the gyratory frequency $\nu_{g}$ is:
\\\\
$\mu = \small\frac{1}{2} e v_{x} r_{B} = \small\frac{1}{2} e v_{x} (v_{x}/\nu_{g}) = \small\frac{1}{2} e c^{2}/2\nu_{g}$
\\\\
$\mu = 1.41061\: 10^{-26}$ J.T$^{-1}$
\\\\
and expressed in terms of the oscillatory frequency $\bar{\nu}_{B}$ it is:
\\\\
$\mu = \eta (\small\frac{1}{2} e v_{x} r_{B}) = \eta [\small\frac{1}{2} e v_{x} (v_{x}/\bar{\nu}_{B})] = \eta (\small\frac{1}{2}\:e c^{2}/2\bar{\nu}_{B})$
\\\\
$\mu = 1.41061 \:10^{-26}$ J.T$^{-1}$
\\\\
In the same way, the reduced Compton oscillatory frequency $\bar{\nu}_{C}$ is:
\\\\
$\bar{\nu}_{C} = c/\bar{\lambda}_{C} = 1.42549 \:10^{24}$ Hz 
\\\\
and the reduced Compton gyratory frequency $\bar{\nu}_{g}$ is:
\\\\
$\bar{\nu}_{g} = c/r_{C} = 2.55203 \:10^{23}$ Hz
\\\\
So, their ratio $\eta_{C} = \nu(oscillatory) / \nu(gyratory)$ is:
\\\\
$\eta_{C} = \nu_{C}/\nu_{g}$ = 5.58569 = g
\\\\
and the proton magnetic moment expressed in terms of the oscillatory frequency $\bar{\nu}_{C}$ is:
\\\\
$\mu = \small\frac{1}{2}\eta e (c/2) \bar{\lambda}_{C} = \small\frac{1}{2} \eta e c^{2}/2\nu_{C} = 1.4106 \:10^{-26}$ J.T$^{-1}$
\\\\
and the magnetic moment  expressed in terms of the gyratory frequency  g is:
\\\\
$\mu = \small\frac{1}{2} e (c/2) r_{C} = \small\frac{1}{2} e c^{2}/2 \nu_{g} = 1.4106 \:10^{-26}$ J.T$^{-1}$
\\\\
Thus, the $\eta$-factor, besides being a structural factor, is also a dynamical factor that expresses as well the relationship between gyratory and oscillatory frequencies, and conforms with the g-factor in both features. 

\subsection{Correlation between magnetic momentum and angular momentum}

The angular momentum can be expressed in terms of the deBroglie radius as well as of the Compton radius:
\\\\
$L_{p} = m_{p} v_{x} r_{B}$
\\\\
$L_{p} = (1.6726 \:10^{-27})(2.11985 \:10^{8})(0.83065 \:10^{-15})$ J.s
\\\\
$L_{p} = 2.94525 \:10^{-34}$ J.s
\\\\
Or equivalently: 
\\\\
$L_{p} = m_{p} v_{y} r_{C} = m_{p} (c/2) r_{C}$
\\\\
$L_{p} = (1.6726 \:10^{-27})(2.9979 \:10^{8} / 2)(1.1747 \:10^{-15})$ J.s
\\\\
$L_{p} = 2.94525 \:10^{-34}$ J.s
\\\\
So, the relationship between the proton angular momentum and its spin is:
\\\\
$L_{p}/S = 2.94525 \:10^{-34} / (1.054572 \:10^{-34} / 2) = 5.585695 = g$ 
\\\\
The spin appears to be the resultant of the angular momentum $L_{p}$ due to its precession. 
The precession angle is:
\\\\
$\Theta = \arctan\sqrt{(g/2)^{2}-1}$ = 69.0191º, and thus:
\\\\
$S = \small\frac{1}{2} L_{p} Cos(\Theta) = \hbar/2$, where: $1/Cos(\Theta) = g/2$. 
\\\\
Thus:
\\\\
$L_{p}/S = L_{p} / [\small\frac{1}{2}\: L_{p} Cos(\Theta)] = 2 /Cos(\Theta) = 5.58569 = g$
\\\\
Therefore the relationship between the proton magnetic moment and angular momentum is:
\\\\
$\eta_{p} = \small\frac{1}{2}(e/m_{p}) L_{p}$  	or equivalently:	 $\eta_{p} = \small\frac{1}{2} g (e/m_{p}) S$ 
\\\\
It thus comes out that when using the angular momentum $L_{p}$ of the proton orbital structure, the expression of its magnetic moment does not need any adjusting factor. The standard formulation of the magnetic moment: $\mu = \small\frac{1}{2} (e/m) S$, does not provide the right value of the proton magnetic moment because it uses the spin instead of the actual angular momentum.

\subsection{Correlation between magnetic moment and mass}

The correlation between magnetic moment and mass derives straightforwardly from the relationship between the radius of the structuring orbital and the coupled deBroglie or Compton wavelength. To the structure radius is associated the magnetic moment and to the wavelength is associated the mass. So, in relation with the deBroglie lengths (radius and wavelength), we get:
\\\\
$\mu = \small\frac{1}{2} e (c/\sqrt{2}) r_{B}$		and equally:   $\mu = \small\frac{1}{2} g e (c/\sqrt{2}) \bar{\lambda}_{B}$ 
\\\\		
$m = \hbar /(\bar{\lambda}_{B} c/\sqrt{2})$		and equivalently:	     $m = g [\hbar /(r_{B} c/\sqrt{2})$
\\\\
or still, in terms of the Compton lengths: 
\\\\
$\mu = \small\frac{1}{2} e (c/2) r_{C}$		and equivalently: 	     $\mu = \small\frac{1}{2} g e (c/2) \bar{\lambda}_{C} $
\\\\		
$m = \hbar /(\bar{\lambda}_{C} c)$ 			and equivalently:	     $m = g [\hbar /(r_{C} c)]$
\\\\
So, once the relationship between the radius and the coupled wavelength has been determined, magnetic moment and mass can be both expressed in function of r or $\bar{\lambda}$ at will, but this should not disguise their different origin, the magnetic moment being generated by the gyratory kinetics of the electric charge and the mass by its oscillatory kinetics, through the ensuing successive acceleration-deceleration.

\section{Conclusions}

The orbital model of the structure of elementary particles has provided a coherent and clear cut apprehension of the relationship between the magnetic moment, to which is associated the radius, and the mass, to which is associated the structure wavelength, and so, it has subsequently promoted a deeper insight to the significance of the g-factor. Besides, the combination of the gyratory and oscillatory motions of the structuring electric charge, along with the precession, leads to a complex orbital that could schematize formal quantum orbitals. So, the semi-classical approach of the structuring orbital provides a comprehensive kinetic origin for the complexity of quantum orbitals.
\\\\
Let us stress that, from the orbital model, mass originates at the Fermi length and not at the Planck length, and is directly associated to the structure wavelength, and hence to the oscillation of the vector electric charge defining the orbital structure. Its associated oscillatory frequency generates and fixes the energy of the structuring orbital and therefore its mass. From its expression $m = \hbar/(\bar{\lambda} c)$, it is easily seen that the structure wavelength $\bar{\lambda}$ determines the mass since $\hbar$ and c are constants. The orbital model of elementary particles assumes that all their properties are directly dependent on the quantum state of their structuring orbital. 
\\\\
Once the relationship between the radius and the wavelength has been determined, the magnetic moment and the mass can be expressed as wished in function of r or $\bar{\lambda}$. But this should not be confusing about the fact that the magnetic moment is created by the gyratory kinetics and the mass by the oscillatory kinetics. However, these two kinetics may obey to some coupling, suggested by the fact that other particles show a preferential coupling through the fine-structure constant $\alpha$.
\\\\
The developed semi-classical approach to the proton g-factor relates it to its structure through its electromagnetic structure wavelength and radius. Since care has been taken to stick as much as possible to the most fundamental physical grounds it provides a much easier comprehensive access than its homologue quantum-mechanics handling. Also, apart from bringing a novel access to the physical comprehension of the field of elementary particles, it sets the origin of mass at the Fermi scale, and relates it to their orbital structure and their Compton or deBroglie wavelengths.
\\\\
From the orbital model, elementary particles can be decomposed into oscillators and gyrators. For mathematical convenience oscillators can be treated as strings. Neutral particles could be then regarded as formed by two bonded strings that may vibrate independently, and each one could simultaneously vibrate at several frequencies. Singly charged particles would proceed from the dipole rupture and consequently would just be single strings. To their vibratory frequency is associated the structure wavelength and thus the mass, and to their gyratory frequency is instead associated their magnetic moment. When dealing with the magnetic moment, elementary particles should instead be seen as orbitals spun by a massless electric charge. When both mass and magnetic moment are considered at once then they should be seen as vibrating orbitals.
\\\\
The orbital model has provided a very simple method to apprehend the origin of the g-factor, relating it to structural parameters, the structure wavelength and the electromagnetic radius, as well as to dynamical parameters such as the gyratory and oscillatory frequencies. The need of the g-factor is simply due to the fact that its quantum mechanical formulation is related to the spin instead of its structure angular momentum. The orbital approach to the proton magnetic moment and its g-factor shows a high coherency and a perfect agreement with basic physics and experimental data. Yet, the same method has been applied to other particles and will also be reported.

\section{Bibliography}

\begin{footnotesize}[1] Hellemans, A., Searching for the spin of the proton, Science (AAAS) - Science and Technology, ISSN: 0036-8075 (1995)
\newline
\newline
[2] Hellemans, A., Quark studies put theorists in a spin, Science (AAAS) - Science and Technology, ISSN: 0036-8075 (1996)
\newline
\newline
[3] Jaffe, R.L., Where does the proton really get its Spin?, Physics Today (Sept. 1995)
\newline
\newline
[4] Peterson, I., Proton-go-round: whence does the proton get its spin?, Science News (Sept 6, 1997) 
\newline
\newline
[5] Aidala C., Solving the proton spin crisis at Phenix, Erice (Sept. 4, 2004)
\newline
\newline
[6] EN'YO, H., Gluons don't explain the spin surprise, Riken Research \textbf{3},11 (Nov. 2008)
\newline
\newline
[7] de Florian D., Sassot R., Stratmann M., Vogelsang M., Phys. Rev. Lett. \textbf{101}, 072001 (2008)
\newline
\newline
[8] Thomas A. W., Interplay of Spin and Orbital Angular Momentum in the Proton, Phys. Rev. Lett. 101, 102003 (2008)
\newline
\newline
[9] de Florian D., Phys. Rev. D \textbf{79}, 114014 (2009)
\newline
\newline
[10] Spin: http://en.wikipedia.org/wiki/Spin
\\
G-factors: http://en.wikipedia.org/wiki/G-factor-Calculation
\\
http://en.wikipedia.org/wiki/G-factor
\\
http://en.wikipedia.org/wiki/G-factor-Electron-orbital-g-factor
\newline
\newline
[11] CODATA values (NIST)
http://physics.nist.gov/cgi-bin/cuu/Value?gp|search-for=proton
\newline
\newline
[12] Sangita Haque, L. Begum, Md. Masud Rana, S. Nasmin Rahman and Md. A. Rahman, Determination of proton size from $\pi^{+}p$ and $\pi^{-}p$ scattering at T($\pi^{±}$) = 277-640 MeV (preprints-sources/2003/IC2003052P.pdf)
\\
http://www.ictp.trieste.it/
\\
\begin{footnotesize}Abstract: The pion-nucleon interaction above the $\Delta(1232)$ resonance and in the region of low-lying pion-nucleon resonances is studied. $\pi± p$ elastic scattering at T($\pi±$) = 277-640 MeV characterized by diffraction maxima and minima has been analyzed through the strong absorption model due to Frahn and Venter. The proton radius is determined from the best fit values of the cut-off angular momentum to be 0.85 Fm with a spread of 0.15 Fm. The higher energy pions scan a lower value while the lower energy pions yield a higher value for the size of the proton. The energy averaged radius of the proton size of 0.85 fm obtained in the present analysis is in excellent agreement with proton charge radius of 0.86 Fm quoted in the literature.\end{footnotesize}
\newline
\newline
[13] G. Sardin, Fundamentals of the Orbital Conception of Elementary Particles and of their Application to the Neutron and Nuclear Structure, Physics Essays \textbf{12}, 2 (1999)
\\
http://www.citebase.org/abstract?id=oai:arXiv.org:hep-ph/0102268
\\
http://arxiv.org/ftp/hep-ph/papers/0102/0102268.pdf
\newline
\newline
[14] G. Sardin, Nature and Quantization of the Proton Mass: An Electromagnetic Model(2005)
\\
http://www.citebase.org/abstract?id=oai:arXiv.org:physics/0512108
\\
http://eprintweb.org/S/authors/physics/sa/Sardin/2\end{footnotesize}

\end{document}